\documentclass[reprint,a4paper]{revtex4-1}

\usepackage{geometry} 
\usepackage{setspace}
\usepackage{amsmath}   
\usepackage{graphicx}   
\usepackage{verbatim}   
\usepackage{color}      
\usepackage{subfigure}  
\usepackage{hyperref}   
\usepackage{float}
\begin{document}
\title{Multi-reference extension to virtual crystal approximation pseudo-potentials}
\begin{abstract}
In this study a computational method of the multi-reference VCA (virtual crystal approximation) pseudo-potential generation is presented. This is an extension of that proposed Ramer and Rappe [J. Phys. Chem. Sol. 61, 315(2000)], the scheme of which is in want of the explicit incorporation of semi-core states. To compensate this drawback, a kind of fine tuning applied to the non-multi-reference VCA pseudo-potential; the form of the pseudo-potential is slightly modified within the cut-off radius in order that  the agreements between the pseudo-potential and all-electron calculations are guaranteed both for semi-core and valence states. The improvement in the present work is validated by atomic and crystalline test calculations for the transferability and the lattice constant estimation.
\end{abstract}
\keywords{Virtual crystal approximation (VCA), pseudo-potential, multi-reference}
\author{Akihito Kikuchi}
\email{kikuchi.akihito@canon.co.jp} 
\affiliation{CANON INC., R$\&$D headquarters, 3-30-2 Shimomaruko, Ohta-Ku, Tokyo,146-08501,Japan}
\maketitle

\section{Introduction}
The purpose of the present article is to show a way to generate a reliable pseudo-potential which is applicable to inhomogeneous or super-lattice crystal structures. 

Frozen core model in pseudo-potential generation often suffers from a kind of inaccuracy where the relaxation in the semi-core state is not negligible, as is notable in transition metals, since, in these elements, owing to the overlapping of the semi-core and the valence states, it is inappropriate to assume the semi-core state as chemically inert one. In these elements, for the more accuracy, the pseudo-potential is generated for the topmost, fully-occupied, semi-core p level, not for the empty valence state. This treatment allows the relaxation of the semi-core state in the simulation, but causes an inevitable inaccuracy in the pseudo-potential result. In the atomic pseudo-potential computation, the semi-core state, lying in the lowest p-level, is in exact agreement with the all-electron result as to the energy and the square norm of the amplitudes of the wave-functions outside the cut-off radius. On the other hand, there is no such an agreement for the empty valence p-level, obtained as the second-lowest p-level, in which the discrepancy between the pseudo-potential and the all-electron computations always exists. In order to correct this drawback, it is necessary to introduce some schemes called multi-reference-pseudo-potential (MRPP)\cite{TME}, so that the calculated properties for the valence and semi-core states by the pseudo-potential and all-electron computation coincide with each other. (In some references, this technique is described as "the explicit incorporation of semi-core states".)

Concerning the crystalline pseudo-potential computation, there is an approximating method named "virtual crystal approximation(VCA)". This is a method to deal with the disordered crystal (and also applicable to the super-lattice structure). To represent the inhomogeneity and to reproduce the averaged property of the disordered crystal, the atomic potentials are, in some ways, mixed in accordance with the composition ratio of the atomic replacement. In the simplest implementation, the pseudo-potential for randomly distributed elements A and B is simply averaged by the composition ratio as $x\times V_A+(1-x)\times V_B$. This approximation is not reliable in some cases. Meanwhile, there are more refined ways in VCA; of which the Ramer-Rappe method \cite{VCA} is the most reliable one; it attains more quantitative crystalline simulations than the simply averaged VCA does. The success of this method will be ascribed to its ability, by which the eigenvalue and the charge distribution of the single atom in VCA model can be adjusted to the averaged all-electron computation in accordance with the composition ratio. However, it still lacks the explicit incorporation of semi-core states. 

Touching these two topics, the present work proposes a method to generate multi-reference crystal VCA pseudo-potential. This is an extension to Ramer-Rappe scheme, toward which a kind of fine-tuning is applied so that the multi-reference accuracy, i.e., the explicit incorporation of semi-core states could be achieved.

\section{Computational Method}
In the following, these notations are used; the two elements to be averaged by VCA are denoted by A(B); the principal and angular quantum numbers are denoted by n(n') and l; the wave-functions, eigenvalues and the potentials are denoted as $\phi_{n(n^\prime),l}^{A(B)}(r)$, $\varepsilon_{n(n^\prime),l}^{A(B)}$ and $V_{A(B)}(r)$. The composition ratio is denoted as $\alpha$ for element A and $\beta$ for B.

The wave-functions in the virtual atom(i.e. the averaged image of randomly distributed two elements A and B in the equivalent atomic site) are computed under following conditions.

\begin{itemize}

\item[[I]] The averaged  nuclear potential ( $\alpha+\beta=1$ here.):
 \begin{eqnarray}
V_{nuc}^{VCA}(r)&=&\alpha V^A_{nuc}(r)+\beta V^B_{nuc}(r)\\\nonumber
&=&\frac{-(\alpha Z^A_{nuc}+\beta Z^B_{nuc})}{r}.
\end{eqnarray}

\item [[II]]The averaged eigenvalue (for the lowest eigenstate):
\begin{equation}
\varepsilon_{1,l}^{VCA}=\alpha \varepsilon^A_{n,l}+\beta \varepsilon^B_{n',l}. 
\end{equation}

\item[[III]] The boundary condition towards the infinity:
\begin{equation}
\phi_{1,l}^{VCA}(r)\rightarrow 0 \quad as\quad r \rightarrow 0.
\end{equation}

\item[[IV]] The norm conserving condition:
\begin{eqnarray}
&&\int_{r_c}^{\infty} \left|\phi_{1,l}^{VCA}(r)\right|^2 r^2 dr	\\\nonumber
  & =&\alpha \int_{r_c}^{\infty}   \left|\phi_{n,l}^A(r)\right|^2 r^2 dr
 +  \beta \int_{r_c}^\infty\left|\phi_{n^\prime,l}^B (r)\right|^2 r^2 dr.		
\end{eqnarray}

\item[[V]] The averaged core charge:
\begin{equation}
\rho_{core}(r)=\alpha\, \rho_{core}^A(r)
+\beta\,\rho_{core}^B (r).		
\end{equation}

\end{itemize}

The computation proceeds as follows.	

\begin{itemize}
\item[1)]With the given energy $\varepsilon_{1,l}^{AVG}$, the node-less numerical solution $\phi_{1,l}^{VCA} (r)$ of the atomic wave equation is calculated by an inward integration from the infinity to the cut-off radius r$_c$. The potential is determined by the conditions of [I] and [V]. The wave-function between the cut-off radius and the infinity is normalized so that the condition [IV] is satisfied.

\item[2)]Prepare the complete pseudo-wave-function. For this purpose, $\phi_{1,l}^{VCA} (r)$ is extended toward the origin (r=0) by some analytic function. The pseudo-charge and the electronic potentials are computed now.
	
\item[3)] Iterate 1) and 2) and obtain the self-consistent charge.

\item[4)] Determine the VCA pseudo-potential which generates $\phi_{1,l}^{VCA} (r)$.

\item[5)] Apply the fine tuning to realize the multi-reference. This step follows Teter's method to generate an extended type of the norm conserving pseudo-potential \cite{TETER}.In this method, the self consistent potential $V_{scf} (r)$ is modified near the origin by means of the cut-off function $h(r)$ and the augmentation terms $\sum_{i=1}^n a_i \,g_i(r)$. The coefficients $a_i$ are adjustable ones so that the computed result will take the required value. (It is noted here that the extended pseudo-potential by Teter is, in origin, not the approach to explicit incorporation of semi-core states. Its purpose is to improve the transferable property of the pseudo-potential by keeping the agreement of the "chemical-hardness" between the pseudo-potential and all-electron computation.)

In the present VCA case for multi-reference extension, following conditions should be satisfied: the eigenvalue and the square norm outside the cut-off radius of the second lowest orbital $\phi_{2,l}^{VCA} (r)$ agree with the averaged all-electron result. These conditions are given as
\begin{equation}
\varepsilon_{2,l}^{AVG}=\alpha \,\varepsilon^A_{n+1,l}+\beta \,\varepsilon^B_{n^\prime+1,l} ,
\label{eq:eigenupper}
\end{equation}

and 

\begin{eqnarray}
&&\int_{r_c}^{\infty} \left|\phi_{2,l}^{VCA}(r)\right|^2 r^2 dr	\\\nonumber
   &=&\alpha \int_{r_c}^{\infty}   \left|\phi_{n+1,l}^A(r)\right|^2 r^2 dr\\\nonumber
   & +&\beta \int_{r_c}^\infty\left|\phi_{n^\prime+1,l}^B (r)\right|^2 r^2 dr.		
    \label{eq:normupper}
\end{eqnarray}

\end{itemize}

The computational steps from 1) to 4) are the same as those in the original Ramer-Rappe method; the step at 5) is the essential extension by the present work. In this step, the fine tuning proceeds in the following way. The conditions to be satisfied are newly given as:

\begin{itemize}
\item[[I']] The pseudo-potential is readjusted from the screened VCA pseudo-potential by means of above cut-off functions and coefficients.In the implementation of the present work, the functional form is given as

\begin{eqnarray}
&&V_{l,ps}^{MRPP} (r)\\\nonumber
&=&\sum_{i=1}^n a_i\,g_i (r)+h(r/r_c )\,V_{l,ps}^{VCA} (r)
+c_l\left(1-h(r/r_c ) \right),
\end{eqnarray}
where $h(r)$ is a cut-off function which is zero at the origin and becomes unity out of the cut-off radius. $g_i (r)$ are functions having i-1 nodes, which are unity at the origin and go to zero with the zero slope at the cut-off radius. This is an extension of the functional form of the provisional pseudo-potential in BHS scheme, presented as eq.(2.10) in ref.\cite{BHS}. Besides, there is a slight altercation from the expression used by Teter\cite{TETER}, to which the extra  term $c_l\left(1-h(r/r_c )\right)$ is included with the adjustable coefficient $c_l$.

\item[[II']] The computed eigenvalue of the lowest state should be the same as [II].

\item[[III']] The boundary condition for orbitals is the same as [III].

\item[[IV']] The norm conserving condition of orbitals is the same as [IV].(Note:this condition is imposed only on the lowest eigenstates. The norm-conserving condition for the second-lowest eigenstate will be ascertained AFTER the fine-tuning.)

\item[[V']] There is no core charge:
\begin{equation}
\rho_{core}(r)=0.
\end{equation}
\end{itemize}

The fine tuning computation takes following steps.

\begin{itemize}
\item[f0)] Set initial values for coefficients $a_i$ and $c_l$.

\item[f1)] Calculate the lowest, node-less numerical solution $\phi_{1,l}^{VCA} (r)$ and adjust the coefficient $c_l$, so that the eigenvalue agrees with $\varepsilon_{1,l}^{AVG}$.

\item[f2)] Renormalise the wave-function $\phi_{1,l}^{VCA} (r)$, so that norm-conservation conservation condition shall be satisfied.  The part of the wave-function must be replaced by some analytic function between the origin and the cut-off radius.

\item[f3)] From $\phi_{1,l}^{VCA} (r)$, the screened pseudo-potential $W_{l,ps}^{VCA} (r)$  is constructed. 

\item[f4)]  Compute the second lowest state $\phi_{2,l}^{VCA} (r)$ in the above-mentioned $W_{l,ps}^{VCA} (r)$. Check whether the conditions in equations (\ref{eq:eigenupper}) and (\ref{eq:normupper}) are satisfied. If not, change the coefficients ${a_i}$ for the cut-off function and start again from f1). In the present work, for the purpose of the optimization, the Powell algorithm without derivatives \cite{POWELL} is applied to reduce the square norm of the residues associated with these equations.

\end{itemize}

Some technical minute points in the implementation of the present work is explained here. 

The provisional VCA pseudo-potential generation before the fine tuning does not always follow the original way proposed by Ramer and Rappe. In the elongation of the inward solution of the atomic equation at the step 2), the present work adopted Troullier-Martins-type analytic function, while Ramer and Rappe used their own. The analytic function used in the present study is the original Troullier-Martins type as given in ref. \cite{TM}, not of the more complicated one as in ref. \cite{TME} for multi-reference extension in the pseudo-potential generation of simple elements. The choice of the Toullier-Martins type function, due to the large number of adjustable variables, sometimes causes the instability in the self-consistent cycle. When confronted to this instability, the Kerker-type analytic function\cite{KERKER}, which contains fewer variables, is more stable. 

The fine tuning process goes in a hybrid way of BHS and Troullier-Martins  scheme. The atomic wave-function is computed under the modified screened-potential in BHS-scheme-like way, from which the node-less solution is obtained. The node-less solution has the necessitated eigenvalue. Then the part of the wave-function is replaced by the Troullier-Martins type analytic function around the nucleus so that the square-norm between the cut-off radius and the infinity agrees with the averaged all-electron value.( For this purpose, the original Troullier-Martins type analytic function is also adopted, not the extended one.) Thus the pseudo-potential, generated by the inversion of the Schrodinger equation, has the "soft-core property" that is characteristic in Troullier-Martins scheme, which shall ensure the good transferability\cite{TM}. The necessitated eigenvalue and the square norm in the valence orbital wave-function(the one-node solution) are attained by the optimization in the augmented form of the screened provisional pseudo-potential. The augmentation and cut-off functions of the self consistent potential ($g_i (r)$and $h(r)$) are those proposed in the appendix of \cite{TETER}, except that the parameters used here are original ones.(These parameters are given in the appendix of this article.)

With regards to the way of the multi-reference pseudo-potential generation, the difference between the method of Reis et al.\cite{TME} and the present work is as follows:  in the former, the realization of multi-reference is achieved by increasing the number of the parameters in the analytic function of the pseudo-orbital between the origin and the cut-off radius.The multi-reference pseudo-potential is computed from the optimized pseudo-orbital. Meanwhile, the present work is the fine-tuning process applied to the form of the provisional pseudo-potential for semi-core electronic configuration. The screened pseudo potential around the nuclei is modified by the addition of the set of analytic functions, the parameters of which are variationally determined so that the  multi-reference extension is obtained. That is to say, the starting points for enforcing the multi-reference extension in the former is the larger number of freedom in the pseudo-orbital, while that in the present work is the more broader freedom assigned to the modification in the screened-pseudo potential.

\section{Result and discussion}

In this section, the examples of the multi-reference pseudo-potential computations by the present work are shown. 

As an example, the multi-reference VCA pseudo-potentials are calculated for several virtual atoms. These are as follows:(1)Ti$_{0.5}$Zr$_{0.5}$ ;(2)Ti$_{0.5}$Hf$_{0.5}$ (3)Ti$_{0.5}$Zr$_{0.25}$Hf$_{0.25}$ ;(4)Ti$_{0.5}$V$_{0.5}$.   The cases (1)-(3) are mixing of elements in the same column in the periodic table. On the other hand, the case (4) is the mixing of elements neighbouring in the same low in the periodic table.   The all-electron results to be averaged are computed  non-relativistically, by means of Ceperlay-Alder type exchange-correlation with Perdew-Zunger parametrization\cite{PW}, at the electron configurations  $p^6s^2d^2$ (for Ti,Zr,Hf) and $p^6s^3d^2$ ( for V); here the p-states are chosen to be semi-core levels. In cases(1)-(3) the cut-off radii are 2.54, 2.96, 2.25 a.u. for s, p ,d orbitals. These values are the same as used in ref.\cite{TIPP}, which are applied in the pseudo-potential generation for Ti with the configuration $4s^24p^03d^2 $. In cases (4),the cut-off radii are 2.10, 2.10, 2.10 a.u. for s, p ,d orbitals. Figure 1 shows the calculated pseudo-potential of the case (1), in which the s,p,d components and the Coulombic potential $-Z_{nuc}/r$ (denoted as "Z-ion" to see the asymptotic behaviour) are illustrated. Table 1 shows the eigenvalues and square norms (computed between the cut-off radius and the infinity) of  the semi-core, or the lowest p states and  the valence, or the second lowest p states. In this table, first, the results by the averaged all-electron computation, secondly, those by the semi-core pseudo-potential (of Ramer-Rappe type,without multi-reference fine-tuning) and,thirdly, those by the present multi-reference method are compared. These results shows that the computation by the present method is able to assure the best agreement between the pseudo-potential and the all-electron calculation in all eigenvalues and square norms, at least in the fixed configuration for which the potential is constructed. The purpose of these calculations are to show the efficacy of the proposed algorithm. Hereafter, for the purpose of the inspection for the property of the multi-reference VCA pseudo-potential, the computations are concentrated to Ti$_{x}$Zr$_{1-x}$, which could be compared with reliable experimental data.

\begin{figure*}
\centering
\includegraphics[scale=0.5,bb=0 0 640 480]{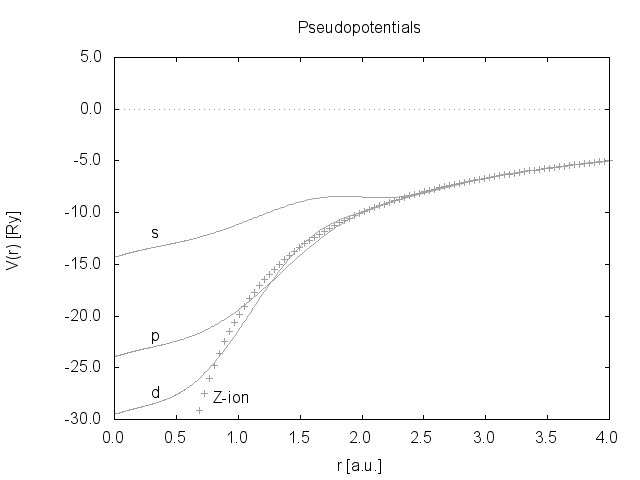}
\caption{This figure shows the multi-reference pseudo-potential for Ti$_{0.5}$Zr$_{0.5}$, computed by the present method. The components of s, p, d channels are shown together with the coulombic potential $-Z_{nuc}/r$.}
\label{fig:001}
\end{figure*}

\begin{table*}
\begin{tabular}{l|c|c|c}
\hline
 (1) Ti$_{0.5}$Zr$_{0.5}$  & All-electron & Ramer-Rappe                & The present method \\ 
 &         & semi-core pseudo-potential &  \\ 
\hline semi-core p eigenvalue  & -2.609 & -2.609  &-2.609  \\ 
\hline semi-core p square norm & 0.993  &  0.993  & 0.993  \\ 
\hline valence p eigenvalue    & -0.115 & -0.124  &-0.115  \\ 
\hline valence p square norm   & 0.873  &  0.860  & 0.873 \\
\hline 
 (2) Ti$_{0.5}$Hf$_{0.5}$ & All-electron & Ramer-Rappe                & The present method \\ 
 &         & semi-core pseudo-potential &  \\ 
\hline semi-core p eigenvalue  & -2.669 & -2.669  &-2.669  \\ 
\hline semi-core p square norm & 0.994  &  0.994  & 0.994  \\ 
\hline valence p eigenvalue    & -0.115 & -0.127  &-0.115  \\ 
\hline valence p square norm   & 0.874  &  0.852  & 0.874 \\
\hline 
 (3) Ti$_{0.5}$Zr$_{0.25}$Hf$_{0.25}$ & All-electron & Ramer-Rappe                & The present method \\ 
 &         & semi-core pseudo-potential &  \\ 
\hline semi-core p eigenvalue  & -2.639 & -2.639  &-2.639  \\ 
\hline semi-core p square norm & 0.995  &  0.995  & 0.995  \\ 
\hline valence p eigenvalue    & -0.115 & -0.125  &-0.115  \\ 
\hline valence p square norm   & 0.868  &  0.851  & 0.868 \\
\hline 
 (4) Ti$_{0.5}$V$_{0.5}$ & All-electron & Ramer-Rappe                & The present method \\ 
 &         & semi-core pseudo-potential &  \\ 
\hline semi-core p eigenvalue  & -3.032 & -3.032  &-3.032  \\ 
\hline semi-core p square norm & 0.974  &  0.974  & 0.974  \\ 
\hline valence p eigenvalue    & -0.114 & -0.118  &-0.114  \\ 
\hline valence p square norm   & 0.958  &  0.954  & 0.958 \\
\hline 

\end{tabular}
\label{tab:4p5peigvals}
\caption{This table shows the computed result of eigenvalues and norm-conserving conditions in several types of  virtual atoms; the comparison between Ramer-Rappe and the present method is given.} 
\end{table*}

As to the transferability of the potential, table 2 shows the results of the configuration test to Ti$_x$Zr$_{(1-x)}$ (x=0.5), in which the eigenvalues in different electron configurations are given. First, the result by averaging all-electron computations, secondly, that by multi-reference VCA pseudo-potential in the present work, thirdly, that by VCA non-multi-reference pseudo-potential (of Ramer-Rappe type), and, finally their differences are listed and compared in columns. These results shows that the present method of multi-reference VCA pseudo-potential assures the transferability of VCA pseudo-potentials more firmly than the computation of simple semi-core pseudo-potential without multi-reference extension does. The simplest VCA, achieved by the simple averaging of the pseudo-potentials shows large errors from all-electron results, especially in d-levels. The more refined ways, those of Ramer-Rappe and the present work afford us more reliable results. As to the improvement by the present work, it is exemplified in the eigenvalues for the valence p-orbital, in which the Ramer-Rappe pseudo-potential without multi-reference shows the larger discrepancy from the all-electron results as the electron configuration varies from the ground one; the computed results show that the growth of such errors is suppressed by the multi-reference VCA of the present work. Meanwhile, as for the other eigenvalues, the discrepancy are of the same order in Ramer-Rappe method and the present method. (For the purpose of checking pseudo-potentials, logarithmic derivatives are computed in usual. That can also be computed by VCA pseudo-potential. But there is no comparable counterpart by the all-electron results; the average of logarithmic derivatives for elements Ti and Zr is meaningless.)

\begin{table*}
  \centering
    \begin{tabular}{l|l|c|l|l|l|l|r|r|r}
    \hline
    Conf. &  & Occupation & I     & II    & III   & IV    & I-III  & II-III & IV-III \\
     No.     &  & number &          &     &     &   &  &
    \\
          &  &    & Ramer         & Present  & All    & Simplest  &  &
    \\
          &  &    & -Rappe         & Work    & Electron    & VCA  &  &  
    \\\hline
    1     & 1s    & 2     & -0.3301  & -0.3301  & -0.3301  & -0.3428  & 0.0\% & 0.0\% & 3.8\% \\
         & 2p    & 6     & -2.6089  & -2.6089  & -2.6089  & -2.5941  & 0.0\% & 0.0\% & -0.6\% \\
         & 3d    & 2     & -0.3205  & -0.3205  & -0.3205  & -0.2506  & 0.0\% & 0.0\% & -21.8\% \\
         & 3p    & 0     & -0.1002  & -0.1153  & -0.1153  & -0.1307  & -13.1\% & 0.0\% & 13.3\% \\
        \hline
    2     & 1s    & 2     & -0.4525  & -0.4410  & -0.4454  & -0.4402  & 1.6\% & -1.0\% & -1.2\% \\
         & 2p    & 6     & -2.9580  & -2.9399  & -2.9762  & -2.8869  & -0.6\% & -1.2\% & -3.0\% \\
         & 3d    & 1     & -0.6020  & -0.5833  & -0.6169  & -0.4636  & -2.4\% & -5.6\% & -24.9\% \\
         & 3p    & 1     & -0.1754  & -0.1900  & -0.1899  & -0.1977  & -7.6\% & 0.1\% & 4.1\% \\\hline
      3   & 1s    & 1     & -0.3940  & -0.3835  & -0.3833  & -0.3903  & 2.8\% & 0.1\% & 1.8\% \\
         & 2p    & 6     & -2.7116  & -2.6935  & -2.6998  & -2.6670  & 0.4\% & -0.2\% & -1.21\% \\
         & 3d    & 2     & -0.4151  & -0.3991  & -0.4025  & -0.3174  & 3.1\% & -0.8\% & -21.1\% \\
         & 3p    & 1     & -0.1492  & -0.1585  & -0.1582  & -0.1691  & -5.7\% & 0.2\% & 6.9\% \\\hline
    4     & 1s    & 2     & -0.8418  & -0.8416  & -0.8483  & -0.8433  & -0.8\% & -0.8\% & -0.6\% \\
         & 2p    & 6     & -3.3969  & -3.3987  & -3.4378  & -3.3477  & -1.2\% & -1.1\% & -2.62\% \\
         & 3d    & 1     & -1.0344  & -1.0336  & -1.0711  & -0.9127  & -3.4\% & -3.5\% & -14.8\% \\
         & 3p    & 0     & -0.5194  & -0.5526  & -0.5542  & -0.5650  & -6.3\% & -0.3\% & 1.9\% \\\hline
    5     & 1s    & 2     & -1.4536  & -1.4530  & -1.4728  & -1.4440  & -1.3\% & -1.4\% & -1.95\% \\
         & 2p    & 6     & -4.3235  & -4.3286  & -4.4509  & -4.2685  & -2.9\% & -2.8\% & -4.10\% \\
         & 3d    & 0     & -1.8885  & -1.8867  & -2.0057  & -1.7306  & -5.8\% & -5.9\% & -13.72\% \\
         & 3p    & 0     & -1.0448  & -1.0957  & -1.1092  & -1.1025  & -5.8\% & -1.2\% & -0.6\% \\\hline
    6     & 1s    & 1     & -0.7524  & -0.7524  & -0.7521  & -0.7649  & 0.04\% & 0.03\% & 1.7\% \\
         & 2p    & 6     & -3.1188  & -3.1185  & -3.1241  & -3.0926  & -0.2\% & -0.2\% & -1.0\% \\
         & 3d    & 2     & -0.8154  & -0.8153  & -0.8191  & -0.7324  & -0.5\% & -0.5\% & -10.59\% \\
         & 3p    & 0     & -0.4585  & -0.4854  & -0.4851  & -0.5055  & -5.5\% & 0.1\% & 4.22\% \\\hline
    7     & 1s    & 0     & -1.2058  & -1.2056  & -1.2001  & -1.2198  & 0.5\% & 0.5\% & 1.7\% \\
         & 2p    & 6     & -3.7126  & -3.7116  & -3.7227  & -3.6741  & -0.3\% & -0.3\% & -1.31\% \\
         & 3d    & 2     & -1.3833  & -1.3830  & -1.3900  & -1.2844  & -0.5\% & -0.5\% & -7.60\% \\
         & 3p    & 0     & -0.8443  & -0.8836  & -0.8802  & -0.9105  & -4.1\% & 0.4\% & 3.5\% \\
    \hline
    \end{tabular}%
\caption{Configuration test for valence and semi-core eigenvalues. In this table, the following types of the calculations are given in each column respectively. Column I: VCA by Ramer-Rappe, semi-core configuration,  no-multi-reference type. Column II: VCA by the present work, multi-reference type.Column III: All electron result (averaged). Column IV: Simplest VCA; this is the simple average of Ti and Zr pseudo-potentials; in which the averaged potentials are multi-reference ones, computed by the present method for simple element(x=0 or 1).The differences between the pseudo-potential and all-electron computations are also given in percentage in  three columns in the right side, denoted as "I-III","II-III","IV-III". The principal quantum number is re-indexed to the eigenstates in the pseudo-potential computation, so that the symbols 2p and 3p signify the semi-core and valence p states respectively.}
\label{CONFTEST2}
\end{table*}

It should be noted here that in the above example calculation, the cut-off radii chosen there are rather large ones. This is not unrealistic choice; the cut-off radii about these values had actually been applied to simulations in transition-metal materials\cite{TIPP}. They will be sufficient for the purpose of the band structure computation. As for the crystal structural properties, however, smaller cut-off radii should be chosen in some cases. As a test calculation, the lattice constant of PbTi$_{0.48}$Zr$_{0.52}$O$_3$ is computed by means of VCA pseudo-potentials generated by the present scheme. The computed lattice constants are given in  table 3 and 4, and compared to the experimental data by Noheda et al. The LDA computation (by Ceperlay-Ader type correlation with Perdew-Zunger parametrization\cite{PW}) are executed with 64 $(4\times4\times4)$ k-points ,and at the plane-wave cut-off energy of 30 Hartrees. The cut-off radii for VCA pseudo-potential for Ti/Zr are taken to be 2.10 a.u. in s orbital and 1.80 a.u. in p and d orbitals. (These cut-off radii are smaller than those in the atomic test computation as is shown in the above.) The pseudo-potentials for Pb and O are generated by Troullier-Martins type and the cut-off radii are taken from former studies; the cut-off radii for Oxygen 2s and 2p orbitals are 1.45 a.u.\cite{TM}; those values for Pb 6s and 6p orbitals are 3.18 a.u.\cite{PBPS}. The core-correction is applied to the Pb core charge, which is included in the evaluation of the exchange-correlation potential in the LDA crystalline computation. The local part in Kleinmen-Bylander form is taken to be s (l=0) component for those three elements. The lattice constants are evaluated by the least-squares fitting to the total energies by the Murnaghan equation; in these computations the relative direction of the crystal axes, the atomic fractional (or decimal) coordinates with respect to the lattice vectors and the ratios of the lengths of the crystal axes (c/a and b/a) are fixed to the experimental results; the one of the length of the lattice axes (a) is the only one variable in the fitting. The results by the present work for the tetragonal and monoclinic PZT show in fairly good agreements with the experiments\cite{NOHEDA}. (The comparable experimental data are given in Table I and III in that reference. The difference from the computation is of the order of $~1\%$.) To illustrate the improvement by multi-reference-extension, the computed result for the tetragonal phase are compared with the two cases of computations, each of which adopted different pseudo-potentials. In the comparable case (1), the pseudo-potential is generated for the empty valence p-orbital and the lower p-levels are regarded as the inactive core; in the case (2), without multi-reference extension, the potential is generated for the fully-occupied semi-core p-orbital.  The cut-off radii of the virtual atom in the case (1) are 2.54, 2.96, 2.25 a.u. for s, p ,d orbitals. Those in the case (2) are the same as in the multi-reference VCA computation. The comparison shows the multi-reference computation by the present work shows better agreement with the experiment than non-multi-reference one, which suggest the improved transferability and the allowance for semi-core relaxation in the virtual atom can yield the more reliable result. It is possible that the more-refined pseudo-potentials (not only for the virtual atom, but, for Pb and O) shall realise more quantitative computations.

\begin{table*}
  \centering
  \begin{tabular}{l|c|c}
    \hline
    Tetragonal phase                  \\
    \hline
    Pseudo-potential type & Configuration &  Lattice constant:a$_t$({\AA}) \\\hline
    Generated for the empty valence p & s$^2$p$^0$d$^2$ &  3.9453   \\
    Generated for semi-core p (not of MRPP) & s$^2$p$^6$d$^2$ &  4.2152   \\
    Present work (of MRPP) & s$^2$p$^6$d$^2$ &  3.9995   \\\hline
    Experiment( Noheda et al.) &       &        4.0460   \\\hline
        \end{tabular}
        
      \label{tab:LAT1}
      \centering
      \caption{Lattice constant calculation (a$_t$) for tetragonal PZT PbTi$_{0.48}$Zr$_{0.52}$O$_3$. }
 \end{table*}
 \begin{table*}     
    \begin{tabular}{l|c|c}
    \hline
    Monoclinic phase                        \\\hline
    Pseudo-potential type & Configuration  & Lattice Constant:a$_m$ ({\AA})  \\\hline
    Present work (of MRPP) & s$^2$p$^6$d$^2$ &  5.6599  \\\hline
    Experiment( Noheda et al.) &               & 5.7220   \\
    \hline
    \end{tabular}
      \caption{Lattice constant (a$_m$ )calculation for monoclinic PZT PbTi$_{0.48}$Zr$_{0.52}$O$_3$.}
  \label{tab:LAT2}
 \end{table*}

As to the crystalline simulation for PZT, in ref.\cite{VCA}, Ramer and Rappe reported the lattice-constant constant calculations for PbZr$_{0.5}$Ti$_{0.5}$O$_3$ by the VCA pseudo-potential without multi-reference extension, in which the computed lattice constants by the VCA agree well with those by the super-lattice model and with the experimental measurement. So there may arise a doubt on the necessity of the multi-reference extension in the VCA proposed in the present work. On this point,however, one should take notice of this: in their VCA computation, Ramer and Rappe used a method of their own, named "designed non-local pseudo-potential(DNL)", in order to enhance the transferability of the pseudo-potentials, in which a flexible augmentation parameter is added to the standard Kleinman-Bylander form of the non-local pseudo-potential\cite{DNL}; furthermore their own technique for optimizing pseudo-potential was also applied\cite{OPTPOT}.(This kind of pseudo-potential construction is not adopted in the present work.) Thus it could be surmised that the reliability of the VCA of Ramer and Rappe will be referable to the synergistic effect of these three techniques, viz., their improvement to VCA, DNL, and their own way of optimized pseudo-potential. In this light, the present work for the multi-reference extension in the VCA can be regarded as an another approach to enhance the pseudo-potential transferability, being as effective and useful as the above-mentioned auxiliary techniques to the VCA.

\section{Concluding remarks}
In this article, the multi-reference extension for VCA pseudo-potentials is developed. The improvement in the transferability of the potentials and the lattice-constant computation are demonstrated by the exemplary calculations of virtual atom Ti/Zr and the virtual crystal Pb(Ti/Zr)O$_3$. This approach is a kind of fine tuning, or, re-making of  the non-multi-reference VCA pseudo-potential into the multi-reference one. It is remarked here that the present work is also the multi-reference extension of the standard pseudo-potential for simple elements, when the composition ratio is set to x=0 or 1. The present work also allows us a broader chance for improving pseudo-potential property; by increasing the numbers of parameters in fine tuning of the pseudo-potential, the other extension, such as, for the conservation of the "chemical hardness", could be easily implemented, in order that the better agreement between pseudo-potential and all-electron computation will be achieved. Accordingly, the present method, if assisted by such extensions, will be of more use to quantitative simulations of disordered crystals.

\appendix
\section*{Appendix}
 The augmentation and cut-off functions in the present work are as follows. 
  
 If $0 \le x\le 1$ then
 $$g_m (x)=\frac{\sin(m\pi x)}{m\pi x} \sum_{j=0}^5 b_{mj} x^{2j}, 
 $$
 $$
 h(x)=1-(1-x^5 )^2,  
 $$
 else
 $$ 
 g_m (x)=0,
 $$
 $$
 h(x)=1,
 $$
 in which coefficients are given by
 $$
 b_{m0}=1,b_{m1}=\frac{m^2 \pi^2}{6},b_{m2}=-10-\frac{m^2 \pi^2}{3},
 $$
 $$
 b_{m3}=20+m^2 \pi^2,b_{m4}=-15-\frac{2m^2 \pi^2}{3},b_{m5}=4+\frac{m^2 \pi^2}{6}.
 $$
 These values assure the following property of the function and the higher derivatives:
$$
  g_m (0)-1=g_m^{(1)} (0)=g_m^{(2) } (0)=g_m^{(3)} (0)=0, g_m^{(4)}(0)\neq 0,
$$
$$  
  g_m (1)=g_m^{(1)} (1)=g_m^{(2)} (1) =g_m^{(3)}(1) =g_m^{(4)} (1)=0.
$$

\end{document}